\documentclass[pra,showpacs,twocolumn,10pt]{revtex4}

\usepackage{amsmath}
\usepackage{dcolumn}
\usepackage[cp1251]{inputenc}
\usepackage[english]{babel}
\usepackage{graphicx}

\begin{document}

\title{Biphoton ququarts as either pure or mixed states, features and reconstruction from coincidence measurements}
\author{M.V. Fedorov}
\affiliation{A.M.Prokhorov General Physics Institute, Russian Academy of Science, Moscow, Russia\\ e-mail: fedorovmv@gmail.com}

\begin{abstract}
Features of biphoton polarization-frequency ququarts are considered. Their wave functions are defined as functions of both polarization and frequency variables of photons with the symmetry obligatory for two-boson states taken into account. In experiments, biphoton ququarts can display different features in dependence on whether experiments involve  purely polarization or (alternatively) polarization-frequency measurements. If in experiments one uses only polarization measurements, the originally pure states of ququarts can be seen as mixed biphoton polarization states. Features of such states are described and discussed in details. Schemes of coincidence measurements for reconstruction of the ququart's parameters are suggested and described.
\end{abstract}
\maketitle

\section{Introduction}

Biphoton polarization-frequency ququarts can be produced in processes of collinear Spontaneous Parametric Down-Conversion (SPDC) non-degenerate with respect to frequencies of photons \cite{Bogd-06}.  In such states  photons have two degrees of freedom: polarization and frequency. In terms of photon polarization and frequency variables, $\sigma$ and $\omega$, each of them can take independently one of two values: $\sigma=H\,{\rm or}\,V$ (horizontal or vertical polarization) and $\omega=\omega_h\,{\rm or}\,\omega_l$ (high or low frequencies) \cite{NJP},\cite{PRA}. For experimental investigation of such states one has to use detectors provided with both polarizers and frequency filters. However, sometimes it is more convenient and, maybe, even more interesting to use only polarizers and wide-band detectors, non-selective in frequencies. In theoretical description, such situation corresponds to averaging of the ququart's states over photon frequencies, or taking traces of the biphoton density matrix with respect to frequency variables of photons. In a general case, this gives rise to  two-qubit biphoton mixed polarization states (MPS) \cite{PRA}. In this paper features of such mixed states are briefly summarized and schemes for measuring their parameters are described. The method to be described is based on a series of coincidence measurements. In comparison with the earlier suggested  general scheme of measuring the ququarts's parameter (Appendix B of Ref. \cite{NJP}) the case of MPS has its own rather interesting peculiarities.

\section{Biphoton ququarts and mixed polarization states}

In a general form, the state vector of an arbitrary polarization-frequency biphoton ququart is given by a superposition of four basis state vectors
\begin{gather}
 \nonumber
 |\Psi^{(4)}\rangle=C_1\,a_{H,h}^\dag a_{H,l}^\dag|0\rangle+
 C_2\,a_{V,h}^\dag a_{V,l}^\dag|0\rangle\\
 \label{qqrt-st-vect}
 +C_3\,a_{H,h}^\dag a_{V,l}^\dag|0\rangle+C_4\,a_{V,h}^\dag a_{H,l}^\dag|0\rangle,
\end{gather}
where $C_i$ are arbitrary complex constants obeying the normalization condition $\sum_i|C_i|^2=1$; $a_{H,h}^\dag,a_{H,l}^\dag,a_{V,h}^\dag,\,{\rm and}\,a_{V,l}^\dag$ are one-photon creation operators for four one-photon modes $\{H,\omega_h\}$, $\{H,\omega_l\}$, $\{V,\omega_h\}$, and $\{V,\omega_l\}$. Superpositions of one-photon states $a_{H,h}^\dag|0\rangle,a_{H,l}^\dag|0\rangle,a_{V,h}^\dag|0\rangle,\,{\rm and}\,a_{V,l}^\dag|0\rangle$ form one-photon qudits with the dimensionality of the one-photon Hilbert space $d=4$. Double population of different one-photon modes corresponds to the basis state vectors in Eq. (\ref{qqrt-st-vect}), like $a_{H,h}^\dag a_{H,l}^\dag|0\rangle$, etc. These basis state vectors,as well as their superpositions, can be considered as describing two-qudit states belonging to the two-photon Hilbert space of the dimensionality $D=d^2=16$.

It's very fruitful to use not only state vectors of biphoton ququarts but also their wave functions. In a general case of arbitrary bipartite states with arbitrary variables of two particles $x_1$ and $x_2$ the bipartite wave function $\Psi(x_1,x_2)$ is defined via the bipartite state vector $|\Psi\rangle$ as $\Psi(x_1,x_2)=\langle x_1,x_2|\Psi\rangle$. One reason why it's important to use wave functions is that in terms of wave functions one can use the simplest definition of entanglement . According to this definition a bipartite state is entangled if its wave function cannot be presented in the form of a product of two single-particle wave functions
\begin{equation}
 \label{ent-definition}
 \Psi(x_1,x_2)\neq\varphi(x_1)\,\chi(x_2).
\end{equation}
Otherwise, if one can find such function $\varphi(x_1)$ and $\chi(x_2)$  that $\Psi(x_1,x_2)=\varphi(x_1)\,\chi(x_2)$, the state is disentangled.

In the case of biphoton states with two degrees of freedom for each photon, for defining two-photon wave functions one has to introduce two pairs of variables, $x_1=\{\sigma_1,\omega_1\}$ and  $x_2=\{\sigma_2,\omega_2\}$. As photons are indistinguishable, we cannot attribute variable numbers 1 and 2 to any of two photons, though we know for sure that the amount of variables equals the amount of photons (2) times the amount of degrees of freedom (2), which gives 4 or two pairs. The biphoton wave function corresponding to the state vector of Eq. (\ref{qqrt-st-vect}) can be found with the help of the general rules of quantum electrodynamics (see, e.g., \cite{Schweber}). The result can be written in different forms. The form most convenient for the further consideration is that related to the use of  polarization and frequency Bell states $\Psi_\pm$:
\begin{gather}
 \nonumber
 \Psi^{(4)}(\sigma_1,\omega_1;\,\sigma_2,\omega_2)=\Psi^{(3)}(\sigma_1,\sigma_2)\Psi_+(\omega_1,\omega_2)\\
 \label{qqrt-wf}
 +B_-\Psi_-(\sigma_1,\sigma_2)\Psi_-(\omega_1,\omega_2),
\end{gather}
where $\Psi^{(3)}(\sigma_1,\sigma_2)$ is the wave function of a purely polarization qutrit
\begin{gather}
 \nonumber
 \Psi^{(3)}(\sigma_1,\sigma_2)=C_1\,\Psi_{HH}(\sigma_1,\sigma_2)+B_+\,\Psi_+(\sigma_1,\sigma_2)\\
 \label{qtr-wf}
 +C_4\,\Psi_{VV}(\sigma_1,\sigma_2)
\end{gather}
with
\begin{gather}
 \label{HH}
 \Psi_{HH}(\sigma_1,\sigma_2)=\delta_{\sigma_1,H}\,\delta_{\sigma_2,H}\equiv \left({1\atop 0}\right)_1^{pol}
 \otimes\left({1\atop 0}\right)_2^{pol},\\
 \label{VV}
 \Psi_{VV}(\sigma_1,\sigma_2)=\delta_{\sigma_1,V}\,\delta_{\sigma_2,V}\equiv \left({0\atop 1}\right)_1^{pol}
 \otimes\left({0\atop 1}\right)_2^{pol},
\end{gather}
and the Bell-state wave functions depending on polarization or frequency variables are given by
\begin{gather}
 \nonumber
 \Psi_\pm(\sigma_1,\sigma_2)=\displaystyle\frac{\delta_{\sigma_1,H}\delta_{\sigma_2,V}
 \pm\delta_{\sigma_1,V}\delta_{\sigma_2,H}}{\sqrt{2}}\\
 \equiv\frac{1}{\sqrt{2}}\left\{\left({1\atop 0}\right)_1^{pol}\otimes\left({0\atop 1}\right)_2^{pol}\pm\left({0\atop 1}\right)_1^{pol}\otimes\left({1\atop 0}\right)_2^{pol}\right\},
 \label{Bell-pol}
\end{gather}
\begin{gather}
 \nonumber
 \Psi_\pm(\omega_1,\omega_2)=\displaystyle\frac{\delta_{\omega_1,\omega_h}\delta_{\omega_2,\omega_l}
 \pm\delta_{\omega_1,\omega_l}\delta_{\omega_2,\omega_h}}{\sqrt{2}} \\
 \equiv\frac{1}{\sqrt{2}}\left\{\left({1\atop 0}\right)_1^{fr}\otimes\left({0\atop 1}\right)_2^{fr}\pm\left({0\atop 1}\right)_1^{fr}\otimes\left({1\atop 0}\right)_2^{fr}\right\};
 \label{Bell-freq}
\end{gather}
superscripts ``$pol"$ and ``$fr"$ in Eqs. (\ref{HH})-(\ref{Bell-freq}) indicate polarization and frequency degrees of freedom. Besides, as it's clear from comparison of the functional and matrix forms of the biphoton wave functions in Eqs. (\ref{HH})-(\ref{Bell-freq}), the upper lines in two-line columns correspond to the horizontal polarization and higher frequency $\omega_h$ and the lower lines - to the vertical polarization and lower frequency $\omega_l$. The constants $B_\pm$ in Eqs. (\ref{qqrt-wf}), (\ref{qtr-wf}) are expressed via $C_{2,3}$ of Eq. (\ref{qqrt-st-vect}) as
\begin{equation}
 \label{B via C}
 B_{\pm}=\frac{C_2\pm C_3}{\sqrt{2}}
\end{equation}
with the normalization condition taking the form $|C_1|^2+|B_+|^2+|B_-|^2+|C_4|^2=1$.

Note that the ququart's wave function (\ref{qqrt-wf}) contains both symmetric and antisymmetric wave functions of the polarization and frequency Bell states. But the antisymmetric Bell-state wave functions $\Psi^{pol}_-$ and $\Psi^{fr}_-$ appear only in the form of their product, which makes the total wave function $\Psi^{(4)}$ symmetric with respect to the transposition of photon variables $1\rightleftharpoons 2$. This is the obligatory feature of two-boson pure states, which often is not taken seriously but which manifests itself, e.g., in the existence of MPS discussed below and in Ref. \cite{PRA}. Besides, owing to the symmetry, both terms in the ququart's wave function (\ref{qqrt-wf}) obey the  entanglement criterion (\ref{ent-definition}) and, hence, all biphoton ququarts are entangled. In some cases this is a purely frequency entanglement  (e.g., if $C_4=B_+=B_-=0$ and $C_1=1$), but in a general case entanglement of biphoton ququarts is an inseparable mixture of the polarization and frequency entanglement.

The density matrix the state (\ref{qqrt-st-vect}),  (\ref{qqrt-wf}) is given by
\begin{equation}
 \label{qqrt-dens-matr}
 \rho^{(4)}=\Psi^{(4)}\otimes\Psi^{(4)\,\dag}.
\end{equation}
This density matrix characterizes pure states. But being averaged over frequency variables, $\rho^{(4)}$ turns into the density matrix of a mixed two-qubit polarization state \cite{PRA}. Written down in the basis $\big\{\Psi_{HH},\,\Psi_+^{pol},\,\Psi_{VV},\,\Psi_-^{pol}\big\}$ and with dropped 12 zero lines and columns , the averaged density matrix takes a rather simple form
\begin{gather}
 \label{rho-mixed}
 \overline{\rho}={\rm Tr}_{\omega_1,\omega_2}\rho^{(4)}=\left(\begin{matrix}\rho^{(3)}& 0\\ 0&|B_-|^2\end{matrix}\right)\\
 \label{pol-4x4}
 =\left(
 \setlength{\extrarowheight}{0.1cm}
 \begin{matrix}
 |C_1|^2 & C_1B_+^* & C_1C_4^* & 0\\
 B_+C_1^* & |B_+|^2 & B_+C_4^* & 0\\
 C_4C_1^* & C_4B_+^* & |C_4|^2 & 0\\
 0 & 0 & 0 & |B_-|^2
 \end{matrix}
 \right),
 \end{gather}
where
\begin{equation}
 \label{rho-qtrt}
 \rho^{(3)}=\Psi^{(3)}\otimes\Psi^{(3)\,\dag}
\end{equation}
is the qutrit's coherence matrix \cite{Klyshko}. In a general case the density matrix $\overline{\rho}$ characterizes a mixed polarization state. The only two exceptions occur in the cases $B_-=0$ and $|B_-|=1$. In the first case the ququart is reduced to qutrit, and in the second case the qutrit's contribution to the quqaurt's wave function (\ref{qqrt-wf}) equals zero (as $C_1=C_4=B_+=0$). In both cases $B_-=0$ and $|B_-|=1$ the ququart's wave function $\Psi^{(4)}$ factorizes for parts depending on the polarization and frequency variables separately. This is the reason why in these cases averaging of a pure polarization-frequency state over frequency variables leaves the remaining polarization state pure. In all other cases ($|B_-|\neq 1,\,0$) there is no factorization for frequency and polarization parts in $\Psi^{(4)}$ and, hence, the state, arising after averaging over frequency variables, is mixed.

There are other forms of presenting the averaged polarization density matrix $\overline{\rho}$ alternative to that of Eq. (\ref{pol-4x4}). One of them used below consists in the presentation of $\overline{\rho}$ in the form of a sum of products of $2\times 2$ single-photon polarization  matrices:
\begin{gather}
 \begin{matrix}
 \overline{\rho}=\left[|C_1|^2\left(\begin{matrix}1&0\\0&0\end{matrix}\right)_1+
 \frac{|B_+|^2+|B_-|^2}{2}\left(\begin{matrix}0&0\\0&1\end{matrix}\right)_1\right]\otimes
 \left(\begin{matrix}1&0\\0&0\end{matrix}
 \right)_2+\\
 \,\\
 \left[\frac{|B_+|^2+|B_-|^2}{2}\left(\begin{matrix}1&0\\0&0\end{matrix}\right)_1
 +|C_4|^2\left(\begin{matrix}0&0\\0&1\end{matrix}\right)_1\right]\otimes
 \left(\begin{matrix}0&0\\0&1\end{matrix}\right)_2+ ...
 \end{matrix}
 \label{pol-2x2x2}
\end{gather}
As it's clear from the definition of $\overline{\rho}$, all matrices in this equation and further below refer to the polarization degree of freedom, with averaging over the frequency variables already performed. For this reason, to shorten notations, here and below we drop the superscript $pol$ common for all arising matrices. In Eq. (\ref{pol-2x2x2}) only four products of $2\times 2$  matrices are shown explicitly. In these four products all matrices are diagonal, whereas in all other 12 products, at least one of the matrices $\left(*\;*\atop{*\;*}\right)_1$ or $\left(*\;*\atop{*\;*}\right)_2$ is off-diagonal. Such terms do not contribute to conditional probabilities analyzed below in section V.

The density matrix of MPS can be further reduced over polarization variables of one of two photons to give rise to the mixed-state reduced density matrix of the form \cite{PRA}
\begin{equation}
 \overline{\rho}_r=\left(
 \begin{matrix}
 |C_1|^2+\frac{|B_+|^2+|B_-|^2}{2} & \frac{C_1B_+^*+B_+C_4^*}{\sqrt{2}}\\
 \frac{C_1^*B_++B_+^*C_4^*}{\sqrt{2}} & |C_4|^2+\frac{|B_+|^2+|B_-|^2}{2}
 \end{matrix} \right).
 \label{rho-red-pol}
\end{equation}

\section{Correlations in mixed biphoton polarization states}

Two correlation parameters found in the general form from the density matrices (\ref{pol-4x4}), (\ref{pol-2x2x2}), (\ref{rho-red-pol}) are the Schmidt parameter $\overline{K}$ and concurrence $\overline{C}$ \cite{PRA}:
\begin{gather}
  \label{K-pol}
  \overline{K}=\frac{2}{1+(1-|B_-|^{\,2})^2-|2C_1C_4-B_+^2|^2}
\end{gather}
and
\begin{equation}
 \label{C-pol}
 \overline{C}=\left||2C_1C_4-B_+^2|-|B_-|^2\right|.
\end{equation}
The concurrence $\overline{C}$ (\ref{C-pol}) characterizes the degree of entanglement or the amount of quantum correlations in MPS. Another quantifier of quantum correlations in such states is the so called relative entropy \cite{Vedral} defined as the ``distance$"$ between the density matrix $\overline{\rho}$ and the density matrix $\sigma$ of the closest disentangled state
\begin{equation}
 \label{S-rel}
 S_{rel}=Tr[\overline{\rho}(\log_2 \rho-\log_2\sigma)].
\end{equation}
For MPS with $C_1=C_4=0$ the relative entropy was found in \cite{PRA} and shown to be less than concurrence at any values of the remaining nonzero parameters $|B_-|$ and $|B_+|=1-|B_-|$. The only exceptions occur at $|B_-|=0,\,1\,{\rm and}\,1/\sqrt{2}$, where the concurrence and relative entropy are equal. Thus, it was found that $S_{rel}\leq {\overline C}$, which can be interpreted as indication that the relative entropy is a better entanglement quantifier than concurrence and that the latter can exaggerate slightly  the degree of entanglement in the case of mixed polarization states. For such states, in accordance with the ideas of Refs. \cite{Vedral,Hamieh,Oh-Kim} one can define the quantifier of classical correlations as the difference between the von Neumann mutual information $I=2S({\overline\rho}_r)-S({\overline\rho})$ and relative entropy $S_{rel}$
\begin{equation}
 \label{C-cl}
 C_{cl}=I-S_{rel}.
\end{equation}

As for the Schmidt parameter of mixed states ${\overline K}$ (\ref{K-pol}), in contrast with pure biphoton polarization states (qutrits), ${\overline K}$ is not related anymore to the concurrence ${\overline C}$ (\ref{C-pol}): ${\overline C}\neq \sqrt{2\left(1-{\overline K}^{\,-1}\right)}$ and {$\overline K\neq 1/(1-{\overline C}^{\,2}/2)$ as in the case of pure states of biphoton qutrits. On the other hand, the Schmidt parameter of MPS remains related to their degree of polarization ${\overline P}$ by the same relation as in the case of pure states of biphoton qutrits
\begin{gather}
 \label{P-K}
 {\overline P}^{\,2}+2\left(1-{\overline K}^{\,-1}\right)=1,
\end{gather}
where ${\overline P}=|{\overline{\vec S}}|$, ${\overline{\vec S}}= Tr\left({\vec\sigma}\,{\overline\rho}_r\right)$ is the vector of Stokes parameters, and
${\vec\sigma}$ is the vector of Pauli matrices. Evidently,
$Tr\left({\vec\sigma}{\overline\rho}_r\,\right)\equiv Tr\left({\vec\sigma}\,\rho^{(4)}\right)$ and, hence, ${\overline P}=P^{(4)}$, i.e., the degree of polarization of the mixed state coincides with the degree of polarization of the original two-qudit ququart, and they both are determined by the Schmidt parameter of the mixed state ${\overline K}$ via Eq. (\ref{P-K}).

As the degree of polarization is a classical concept, we can deduce from Eq. (\ref{P-K}) that in the case of mixed states the Schmidt parameter $\overline{K}$ is related to the amount of classical rather than quantum correlations. In terms of ${\overline K}$, a new  parameter characterizing the amount of classical correlations can be defined as
\begin{equation}
 \label{C-cl-via-K}
 {\overline C_{cl}}=\sqrt{2\left(1-{\overline K}^{\,-1}\right)}.
\end{equation}
It may be interesting to notice that for the state with $C_1=C_4=0$ this parameter coincides with that of (\ref{C-cl})
\begin{equation}
 \label{C-cl=C-cl-K}
 {\overline C_{cl}}=C_{cl}.
\end{equation}
Note finally that in other special cases, $B_-=0$ or $|B_-|=1$, when states averaged over frequencies remain pure, all discussed parameters of quantum and classical correlations coincide with each other and are equal to a half of the von Neumann mutual information
\begin{equation}
 \label{all-coinciduing}
 {\overline C}=S_{rel}=C_{cl}={\overline C}_{cl}=I/2.
\end{equation}
In these cases the relation ${\overline C}=\sqrt{2\left(1-{\overline K}^{\,-1}\right)}$ becomes valid again, and this is the reason why in pure bipartite states the Schmidt parameter $K$ can be used for characterization of the amounts of both quantum and classical correlations.

\section{Comparison with a two-qubit pure-state model of biphoton ququarts}

A picture of mixed two-qubit polarization states described above differs significantly from traditionally widely used model of two-qubit pure-state ququarts. This model starts from the same state vector as given by Eq. (\ref{qqrt-st-vect}). But then frequencies of SPDC photons $\omega_1$ and $\omega_2$ are considered as given numbers rather than variables, e.g., as $\omega_1\equiv\omega_h$ and $\omega_2\equiv\omega_l$. Owing to this, two photons of SPDC pairs are considered as ``partially distinguishable$"$, owing to which the polarization biphoton wave function appears to be not necessarily symmetric with respect to the transposition of particle's variables $1\rightleftharpoons 2$, and can be written in the form
\begin{gather}
 \nonumber
 \Psi^{(4)}_{2\,qb}(\sigma_1, \sigma_2)= C_1\Psi_{HH}(\sigma_1, \sigma_2) +B_+\Psi_+(\sigma_1, \sigma_2)\\
 \label{Psi 2 qb}
+C_4\Psi_{VV}(\sigma_1, \sigma_2)+B_-\Psi_-(\sigma_1, \sigma_2).
\end{gather}
This is a wave function of a pure two-qubit state, and it yields the well known results for the Schmidt parameter and concurrence:
\begin{gather}
 \label{K-2-qb}
 K_{2\,qb}^{(4)}=\frac{2}{2-\left|2C_1C_4-B_+^2+B_-^2\right|^2},\\
 \label{conc-2-qb}
 C_{2\,qb}^{(4)}=\sqrt{2\left(1-K_{2\,qb}^{-1}\right)}=\left|2C_1C_4-B_+^2+B_-^2\right|.
\end{gather}
In a general case, these expressions differ from $\overline{K}$ and $\overline{C}$ of Eqs. (\ref{K-pol}) and (\ref{C-pol}). We believe that the correct results are those based on the picture of two-qudit polarization-frequency bipohoton ququarts and of MPS arising after averaging over frequencies, i.e., the results determined by Eqs. (\ref{K-pol})-(\ref{P-K}). Weak points of the two-qubit theory of biphoton ququarts are evident. Photons of SPDC pairs are always indistinguishable. If there is something that looks like "partial distinguishability", this is an indication that there is, in fact, an additional degree of freedom, and with this degree of freedom taken into account accurately, photons are evidently indistinguishable. Wave functions of two photons in a pure state cannot be asymmetric with respect to the transposition of their variables. Its symmetry is dictated by the Bose-Einstein statistics of photons. This feature is clearly violated in the two-qubit wave function of Eq. (\ref{Psi 2 qb}) where the symmetric and antisymmetric Bell-state wave functions are summed on equal terms. Note however, that a simple symmetrization of the expression in Eq. (\ref{Psi 2 qb}) vanishes the term, proportional to $\Psi_-$, and reduces the ququart's wave function $\Psi_{2qb}^{(4)}$ to that of a qutrit $\Psi^{(3)}$ (\ref{qtr-wf}). To get a correct quqaurt's wave function (\ref{qqrt-wf}), in addition to symmetrization, one has to give freedom to photon frequencies $\omega_{1,2}$ by considering them as variables which can take one of two values each: either $\omega_1=\omega_h$ and $\omega_2=\omega_l$ or $\omega_1=\omega_l$ and $\omega_2=\omega_h$. Actually, this means that we never know which photon has which frequency. Averaging of states of bophoton ququarts over frequencies $\omega_{1,2}$ gives rise to MPS considered here and in Ref. \cite{PRA}. There is no way to get such states in a two-qubit model. In principle, differences between predictions of the theory of mixed states and of the two-qubit model of ququarts can be seen in experiments on measurement of the degree of polarization of biphoton polarization-frequency ququarts. Some simple examples of experimental schemes where these differences are well pronounced are described in Ref. \cite{PRA}.

\section{Reconstruction of ququart's parameters in experiments}

The next questions are how to measure in experiments parameters of MPS and of pure states of polarization-frequency ququarts. It was shown earlier \cite{NJP} that, in principle, a series of coincidence polarization-frequency measurements in three different bases provides sufficient amount of data to get a complete set of equations for finding all ququart's parameters. But equations obtained in such a way were rather complicated and not convenient for practical purposes. Here we will consider first a simpler problem of finding parameters of the above discussed MPS related to biphoton ququarts. And then, at the last stage, we will show how this procedure can be prolonged in a very simple way to reconstruct explicitly all ququart's parameters. Note also that the methods of using series of coincidence measurements for reconstructing parameters of quantum states are alternative to standard and rather widely used methods of quantum tomography for biphoton ququarts \cite {Bogd-03,Bogd-04,Bogd-06}.

\subsection{Independent constants characterizing biphoton qutrits, ququarts and mixed polarization states}
Pure states of qutrits (\ref{qtr-wf}) and ququarts (\ref{qqrt-wf}) are characterized, correspondingly, by three and four complex parameters, $\{C_1,\,B_+,\,C_4 \}$ and $\{C_1,\,B_+,\,C_4,\,B_- \}$, which corresponds to 6 and 8 real constants. But these parameters are not completely independent: there are normalization conditions and, besides, in both cases the common phases of wave functions do not affect measurable quantities and, hence, can be taken having arbitrary most conveniently chosen  given values. These conditions reduce the amount independent real constant parameters characterizing qutrits and ququarts, correspondingly, to 4 and 6. MPS considered above occupy an intermediate position between qutrits and ququarts. Parameters characterizing these states are the same as in the case of pure-state ququarts, $\{C_1,\,B_+,\,C_4,\,B_- \}$. But, in addition to the normalization and common phase conditions we have now one condition more: as seen well from the structure of the density matrix $\overline{\rho}$ written in the form (\ref{pol-4x4}), features of mixed states do not depend of the phase of the parameter $B_-$. This is seen well from the derived expressions for the Schmidt parameter and concurrence (\ref{K-pol}) and (\ref{C-pol}), which depend on $B_-$ only as on $|B_-|$. Also, as seen well from the definition of $\overline{\rho}$ in the form of Eqs. (\ref{rho-mixed}), (\ref{rho-qtrt}), the density matrix of MPS does not depend on the phase of  the qutrit's wave function $\Psi^{(3)}$, which enters into the definition of $\overline{\rho}$ as determining one of its components. Owing to this, the phase of $\Psi^{(3)}$ can be chosen, e.g., in a way, making the parameter $B_+$ real and positive. Thus, we find that in this case MPS are characterized completely by 5 independent real constants: $|C_1|,\,\varphi_1,\,B_+,\,|C_4|, \varphi_4$, where $\varphi_{1,4}$ are phases of the parameters $C_{1,4}$ with the constant $|B_-|$ to be found from the normalization condition.

\subsection{Conditional probabilities and coincidence measurements}

By definition, the conditional probability $\left.w_\sigma\right|_{\sigma^\prime}$ is the probability for a photon 1 to have polarization $\sigma$ under the condition that the second photon (2) of the same pair has polarization $\sigma^\prime$. Relations between the conditional probability and parameters of MPS follow directly from the presentation of the density matrix $\overline{\rho}$ in the form of a sum of products of $2\times 2$ single photon matrices (\ref{pol-2x2x2}):
\begin{gather}
 \label{cond-HH-VV}
 \left.w_H\right|_H=|C_1|^2,\,\left.w_V\right|_V=|C_4|^2,\\
 \label{cond-HV-VH}
 \left.w_H\right|_V=\left.w_V\right|_H=\frac{|B_+|^2+|B_-|^2}{2}.
 \end{gather}
Owing to normalization, $\sum_{\sigma,\sigma^\prime}\left.w_\sigma\right|_{\sigma^\prime}=1$.

In experiment, conditional probabilities can be found from coincidence measurements.
For this goal a biphoton beam has to be divided for two channels 1 and 2 by a non-selective Beam Splitter (BS).
Half of SPDC pairs will be divided between channels, whereas another half of undivided pairs will appear either
in the channel 1 or 2 (see a scheme in Fig. \ref{Fig1}).
\begin{figure}[h]
\centering\includegraphics[width=8.5cm]{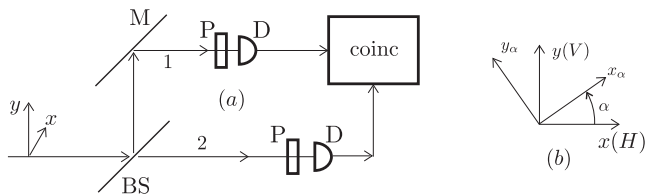}
\caption{{\protect\footnotesize {(a): a scheme of experiment for coincidence measurements, $BS$ - beam splitter, $P$ - polarizers, $D$ - detectors, $M$ - mirror; (b): the horizontal-vertical coordinate frame and the frame turned of an angle $\alpha$. }}}\label{Fig1}
\end{figure}
In coincidence measurements only divided pairs are registered, and for such pairs numbers of channels 1 and 2 can be associated with photon or variable numbers 1 and 2 in Eqs. (\ref{qqrt-wf})-(\ref{Bell-freq}) and (\ref{pol-2x2x2}). Measurements consist in counting photons with a given polarization. Selection of polarization is provided by polarizers installed in each channel in front of detectors. Orientation of each polarizer can be changed independently from horizontal to vertical and vice versa. Each series of measurements with given polarizer orientations has to be performed under identical conditions and to take the same time. As the undivided pairs do not participate in such measurements, the amount of photons with any given polarization accessible for registration is twice less than in the original beam. Besides detectors have some efficiency less than $100\%$, which further diminishes amounts of registered photons. But for relative amounts of registered photons all these losses do not matter. A computer obtaining signals from both detectors registers only coinciding incoming signals and provides measurements of the relative amounts of counts coinciding with conditional probabilities (\ref{cond-HH-VV}), (\ref{cond-HV-VH}). Let $\left.N_\sigma\right|_{\sigma^\prime}$ be the counted number of pairs with the photon polarization $\sigma$ in the channel 1 and $\sigma^\prime$ in the channel 2. Then we have
\begin{equation}
 \label{coinc-cond}
 \frac{\left.N_\sigma\right|_{\sigma^\prime}}{\sum_{\sigma,\sigma^\prime}\left.N_\sigma\right|_{\sigma^\prime}}
 =\left.w_\sigma\right|_{\sigma^\prime}.
\end{equation}
Now Eqs. (\ref{cond-HH-VV}) and (\ref{coinc-cond}) can be used for finding two real constants characterizing MPS, $|C_1|^2$ and $|C_4|^2$, directly from experimental measurements with polarizer in both channels 1 and 2 oriented either horizontally or vertically. Measurements with other orientations $(HV)$ and $(VH)$ are necessary too but only for determination of the normalizing factor in the denominator of Eq. (\ref{coinc-cond}). The sum $|B_+|^2+|B_-|^2=1-|C_1|^2-|C_4|^2$ is determined by normalization but the  constants $B_+$ and $|B_-|^2$ themselves remain undefined, as well as the phases $\varphi_1$ and $\varphi_4$. For finding them other measurements are needed.

\subsection{Conditional probabilities in rotated frames}

Additional information about parameters of MPS can be obtained from coincidence measurements with polarizers in both channels 1 and 2  turned for the same angle $\alpha$ with respect the horizontal or vertical axes.
The results of such measurements (photon counting) are related to the corresponding conditional probabilities by the same relation as in the case $\alpha=0$ (\ref{coinc-cond}):
\begin{equation}
 \label{coinc-cond-alpha}
 \displaystyle\frac{\left.N_\alpha\right|_{\alpha^\prime}}{N_{tot}}=\left.w_\alpha\right|_{\alpha^\prime},\,;
 \displaystyle\frac{\left.N_{90^\circ+\alpha}\right|_{\alpha^\prime}}{N_{tot}}
 =\left.w_{90^\circ+\alpha}\right|_{\alpha^\prime},
 \end{equation}
 where $\alpha^\prime=\alpha$ or $\alpha+90^\circ$ and $N_{tot}= \sum_{\alpha^\prime}[\left.N_\alpha\right|_{\alpha^\prime}
 +\left.N_{90^\circ+\alpha}\right|_{\alpha^\prime}]$

For finding the conditional probabilities $\left.w_\alpha\right|_{\alpha^\prime}$ and $\left.w_{90^\circ+\alpha}\right|_{\alpha^\prime}$, we have to rewrite the same ququart's wave function as given by Eq. (\ref{qqrt-wf}) in the frame turned for an angle $\alpha$ around the  photon propagation axis $Oz$ (in the ($x_\alpha,y_\alpha$)-plane  in Fig. 1$b$). Transformation to this frame is provided by the basic transformation formulas for one-photon states
\begin{gather}
 \label{trasnsf}
 \setlength{\extrarowheight}{0.3cm}
 \begin{matrix}
 \displaystyle\left({1\atop 0}\right)=\cos\alpha\left({1\atop 0}\right)^\alpha-\sin\alpha\left({0\atop 1}\right)^\alpha,\\
 \displaystyle\left({0\atop 1}\right)=\sin\alpha\left({1\atop 0}\right)^\alpha+\cos\alpha\left({0\atop 1}\right)^\alpha,
 \end{matrix}
\end{gather}
where $\left({1\atop 0}\right)^\alpha$ and $\left({0\atop 1}\right)^\alpha$ correspond to polarizations along directions $\alpha$ and $90^\circ+\alpha$. Evidently, the transformations (\ref{trasnsf}) do not affect the frequency part of the ququart's wave function (\ref{qqrt-wf}). Besides, as can be easily checked, the polarization antisymmetric Bell-state wave function is invariant with respect to the transformations (\ref{trasnsf}), i.e., $\Psi_-^\alpha$ expressed via $\left({1\atop 0}\right)^\alpha$ and $\left({0\atop 1}\right)^\alpha$ has the same form as $\Psi_-$ expressed via $\left({1\atop 0}\right)$ and $\left({0\atop 1}\right)$ [Eq. (\ref{Bell-pol})]. This means that after the transformation (\ref{trasnsf}) the part of the ququart's wave function with the product of antisymmetric Bell states does not mix up with the symmetric part, and the form of ququart's wave function is invariant with respect to transformation:
\begin{equation}
 \Psi^{(4)}=\Psi^{(3)\,\alpha}\Psi_+^{fr}+B_-\Psi_-^{pol\, \alpha}\Psi_-^{fr},
 \label{qqrt-wf-alpha}
\end{equation}
where $\Psi^{(3)\,\alpha}$ has the same form in the transformed basis $\left({1\atop 0}\right)^\alpha$ and $\left({0\atop 1}\right)^\alpha$  as $\Psi^{(3)}$ in the original $HV$ basis (\ref{qtr-wf})
\begin{equation}
 \label{qtr-wf-alpha}
 \Psi^{(3)\,\alpha}=C_1^\alpha\,\Psi_{\alpha,\alpha}+B_+^\alpha\,\Psi_+^\alpha
  +C_4^\alpha\,\Psi_{90^\circ+\alpha,90^\circ+\alpha}.
\end{equation}
The coefficients $C_1^\alpha$, $B_+^\alpha$ and $C_4^\alpha$ are easily found to be given by \cite{NJP} [Eq. (A.13)]
\begin{eqnarray}
 \label{C1-alpha}
 C_1^\alpha=\cos^2\alpha \,C_1+\sqrt{2}\cos\alpha\sin\alpha \,B_++\sin^2\alpha \,C_4,\\
 \label{B+-alpha}
 B_+^\alpha=-\sqrt{2}\cos\alpha\sin\alpha\, (C_1-C_4)+\cos 2\alpha \,B_+,\\
  \label{C4-alpha}
 C_4^\alpha=\sin^2\alpha \,C_1-\sqrt{2}\cos\alpha\sin\alpha\,B_++\cos^2\alpha\, C_4,
\end{eqnarray}
The conditional probabilities in the turned frame are defined similarly to their definition in the $HV$-frame (\ref{cond-HH-VV}):
\begin{gather}
 \nonumber
 \left.w_\alpha\right|_\alpha=|C_1^\alpha|^2\\
 \label{w-alpha-alpha}
 =|\cos^2\alpha \,C_1+\sqrt{2}\cos\alpha\sin\alpha \,B_++\sin^2\alpha \,C_4|^2,\\
 \nonumber
 \left.w_{90^\circ+\alpha}\right|_{90^\circ+\alpha}=|C_4^\alpha|^2\\
 \label{w-alpha+90-alpha+90}
 =|\sin^2\alpha \,C_1-\sqrt{2}\cos\alpha\sin\alpha\,B_++\cos^2\alpha\, C_4|^2,\\
 \left.w_{\alpha}\right|_{90^\circ+\alpha}=\left.w_{90^\circ+\alpha}\right|_{\alpha}
 =\frac{|B_+^\alpha|^2+|B_-|^2}{2}.
 \label{w-alpha-alpha+90}
\end{gather}
If $\alpha$ is small, in the linear approximation in $\alpha$, Eqs. (\ref{w-alpha-alpha}) and (\ref{w-alpha+90-alpha+90}) take the form
\begin{gather}
 \label{cos-phi1}
 \left.w_\alpha\right|_\alpha=\left.w_H\right|_H+2\sqrt{2}\,\alpha|C_1|B_+\cos\varphi_1,\\
 \label{cos-phi4}
 \left.w_{90^\circ+\alpha}\right|_{90^\circ+\alpha}=\left.w_V\right|_V-2\sqrt{2}\,\alpha|C_4|B_+\cos\varphi_4,
\end{gather}
where, as assumed, $B_+$ is taken real and positive. By denoting
\begin{equation}
 \label{theta1-theta4}
 \setlength{\extrarowheight}{0.3cm}
 \begin{matrix}
 \lim_{\alpha\rightarrow 0}\frac{\left.w_\alpha\right|_\alpha-\left.w_H\right|_H}{\alpha}=\tan\theta_1,\\
 \lim_{\alpha\rightarrow 0}\frac{\left.w_{90^\circ+\alpha}\right|_{90^\circ+\alpha}-\left.w_V\right|_V}{\alpha}=\tan\theta_4
 \end{matrix},
\end{equation}
we can rewrite Eqs. (\ref{cos-phi1}) and (\ref{cos-phi4}) as
\begin{equation}
 \label{phi via theta}
 \cos\varphi_1=\frac{\tan\theta_1}{2^{3/2}|C_1|B_+},\;
 \cos\varphi_4=\frac{\tan\theta_4}{2^{3/2}|C_4|B_+}.
\end{equation}
Here expressions for the parameters $|C_1|$ and $|C_4|$ in terms of conditional probabilities are known (\ref{cond-HH-VV}), and the derived Eqs. (\ref{cos-phi1}) and (\ref{cos-phi4}) determine directly phases $\varphi_1$ and $\varphi_4$ as functions of $B_+$. For finding $B_+$ [and then $|B_-|$ from Eq. (\ref{cond-HV-VH})] one has to make one measurement and one derivation more, for example, in the frame with $\alpha=45^\circ$, i.e., in the frame turned for $45^\circ$ with respect to the original $HV$- ($xy$-) frame. In this case we find from Eq. (\ref{B+-alpha})  that $B_+^{45^\circ}=|C_1-C_4|^2/2$ and, hence, Eqs. (\ref{cond-HV-VH}) and (\ref{w-alpha-alpha+90}) yield the following equation:
\begin{equation}
 \label{Eq for B+}
 \left.w_{45^\circ}\right|_{135^\circ}=\frac{|C_1-C_4|^2}{4}+\left.w_H\right|_V-\frac{B_+^2}{2},
\end{equation}
where
\begin{equation}
 \label{(C1-C4)2}
 |C_1-C_4|^2=|C_1|^2+|C_4|^2-2|C_1||C_4|\cos (\varphi_1-\varphi_4)
\end{equation}
and
\begin{gather}
 \nonumber
 \cos (\varphi_1-\varphi_4)=\frac{\tan\theta_1}{2^{3/2}|C_1|B_+}\,\frac{\tan\theta_4}{2^{3/2}|C_4|B_+}\\
 \label{cos phi1-phi4}
 +\left[1-\left(\frac{\tan\theta_1}{2^{3/2}|C_1|B_+}\right)^2
 \left(\frac{\tan\theta_4}{2^{3/2}|C_4|B_+}\right)^2\right]^{1/2}.
\end{gather}
With known $C_{1,4}$ and $\tan\varphi_{1,4}$, the only unknown parameter in Eq. (\ref{Eq for B+}) [combined with Eqs. (\ref{(C1-C4)2}) and (\ref{cos phi1-phi4})] is $B_+$,  and this equation has to be solved numerically. When $B_+$ is found, Eq. (\ref{cond-HV-VH}) yields $|B_-|=\left[2\left.w_H\right|_V-|B_+|^2\right]^{1/2}.$ This concludes determination of all five parameters of MPS in terms of conditional probabilities.

\subsection{Scenarios for experimental measurement of the parameters of mixed polarization states}

In all cases, the first step consists in performing at least three series of coincidence measurements of photon numbers with polarizers in channels 1 and 2 installed along either horizontal or vertical directions,
$\left.N_H\right|_H$, $\left.N_H\right|_V$, and $\left.N_V\right|_V$ (because of photon indistinguishability
$\left.N_V\right|_H=\left.N_H\right|_V$). Then, with the help of Eqs. (\ref{cond-HH-VV}) we find two
constants, $|C_1|$ and $|C_4|$ plus the relation between $|B_+|^2$ and $|B_-|^2$ (\ref{cond-HV-VH}). The next steps are different for the situations of zero or non-zero obtained values of the parameters $|C_1|$ and $|C_4|$.

\subsubsection{$C_1=C_4=0$}

If the above-described measurements with horizontal-vertical orientations of polarizers give $|C_1|=|C_4|=0$, the remaining two constants to be found are $|B_+|$ and $|B_-|$, and their measurement is very simple. E.g., from of Eq. (\ref{C1-alpha}) we find $C_1^\alpha=\sin 2\alpha B_+/\sqrt{2}$. Then Eq. (\ref{w-alpha-alpha}) yields
\begin{equation}
 \label{zero C1 C4}
 |B_+|^2=\frac{2\left.w_\alpha\right|_\alpha}{\sin^22\alpha}
 =\frac{2\left.N_\alpha\right|_\alpha/N_{tot}}{\sin^22\alpha}
\end{equation}
and $|B_-|=\sqrt{1-|B_+|^2}.$ Thus, if a complete set of horizontal-vertical coincidence measurements gives $C_1=C_4=0$, for finding $|B_+|$ and $|B_-|$ one has to make only one coincidence measurement more, with identically oriented polarizers $P_1$ and $P_2$ and with arbitrary chosen angle of their orientation $\alpha\neq 0,\,\pi$.

\subsubsection{$C_1\neq 0,\,C_4=0$}

If the horizontal-vertical coincidence measurements give $C_1\neq 0,\,C_4=0$, at $\alpha=45^\circ$ the formula in  Eq. (\ref{B+-alpha}) is reduced to $B_+^{45^\circ}=C_1/\sqrt{2}$, owing to which Eq. (\ref{w-alpha-alpha+90}) takes the form
\begin{equation}
 \label{45-135}
 \left.w_{135^\circ}\right|_{45^\circ}\equiv\frac{\left.N_{135^\circ}\right|_{45^\circ}}{N_{tot}}=
 \frac{|C_1|^2}{4}+\frac{|B_-|^2}{2}.
\end{equation}
This is the equation for finding $|B_-|$, after which $|B_+|$ is also easily found from normalization $|B_-|=\sqrt{1-|C_1|^2-|B_+|^2}$. Thus, for measuring absolute values of all three constants, $|C_1|$, $|B_+|$, and $|B_-|$ ( with $C_4=0$), it is sufficient to complete horizontal-vertical coincidence measurements by measurement of the coincidence number of photons with polarizers $P_1$ and $P_2$ turned, correspondingly, for $135^\circ$ and $45^\circ$ with respect to the horizontal direction. In principle, in addition to these three constants there is one constant more characterizing MPS with $C_4=0$, the phase $\varphi_1$ of the parameter $C_1$ ( with real $B_+$). The way of its measurement is described in the following subsubsection. But it should be noted that in the case $C_4=0$ the MPS correlation parameters $\overline{K}$ (\ref{K-pol}) and $\overline{C}$ (\ref{C-pol}) do not depend on $\varphi_1$. Note also that the described scheme of measurements is valid also in the case $C_1=0,\,C_4\neq 0$.

\subsubsection{Nonzero $C_{1,4}$}

Let now both constants $|C_1|$ and $|C_4|$ found from the horizontal-vertical coincidence measurements be different from zero, $C_{1,4}\neq 0$. Then, the procedure of measuring other MPS parameters is more complicated. In particular, for measuring phases $\varphi_{1,4}$ of $C_{1,4}$, we suggest to use their relations (\ref{phi via theta}) with the parameters $\tan\theta_{1,4}$ (\ref{theta1-theta4}) characterizing the behavior of the function  $\left.w_\alpha\right|_\alpha(\alpha)$ in small vicinities of the points $\alpha=0$ and $\alpha=90^\circ$.
Specifically, for finding $\theta_1$ and $\varphi_1$ we suggest to measure coincidence numbers of photons $\left.N_{\pm\alpha_0}\right|_{\pm\alpha_0}$ with both polarizers $P_1$ and $P_2$ turned for some small angles $\alpha_0$ and $-\alpha_0$ with respect to the horizontal direction (e. g., with $\alpha_0=5^\circ=0.087\,rad\ll 1$). In accordance with Eqs. (\ref{coinc-cond-alpha}) and together with the earlier made measurement of $\left.w_H\right|_H$ this gives three values of the function $\left.w_\alpha\right|_\alpha(\alpha)$ at three values of $\alpha$, $\alpha=-\alpha_0,\,0,\,{\rm and}\,\alpha_0$. In Fig. \ref{Fig2} these values are indicated by letters
\begin{figure}[h]
\centering\includegraphics[width=3.2cm]{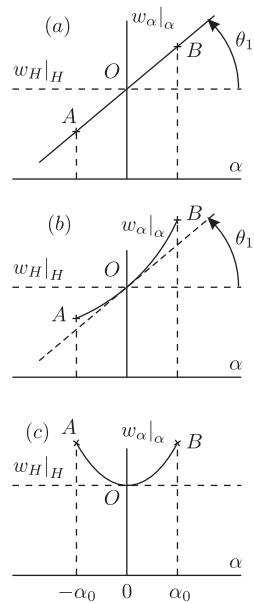}
\caption{{\protect\footnotesize {Conditional probability $\left.w_\alpha\right|_\alpha(\alpha)$ at $|\alpha|\ll 1$} (solid lines) and results of its measurements $A$, $O$, and $B$.}}\label{Fig2}
\end{figure}
$A$, $O$, and $B$. With this three values we can reconstruct the function $\left.w_\alpha\right|_\alpha(\alpha)$ in a small vicinity of the point $\alpha=0$. The pictures $(a),\,(b),\,{\rm and}\,(c)$ correspond to three different possible locations of the points $A$ and $B$. The picture ${(a)}$ corresponds to the case when the points $A$ and $B$ are symmetric with respect to $O$, and all three points $A$, $O$, and $B$ can be connected by a single straight line. In this case the angle $\alpha_0$ is small enough for validity of the approximation linear in $\alpha$ in Eq. (\ref{cos-phi1}) in all range $[-\alpha_0,\alpha_0]$. In the case $(b)$ positions of the points $A$ and $B$ are asymmetric, and the points $A$, $O$, and $B$ can be connected only by a parabola. In this case, the straight line corresponding to the linear approximation of Eq. (\ref{cos-phi1}) is tangent to the parabola in the point $O$. The angle $\theta_1$ is determined in these two cases as the angle between the horizontal line and either the line $\left.w_\alpha\right|_\alpha(\alpha)$ in the case $(a)$ or the line tangent to the curve $\left.w_\alpha\right|_\alpha(\alpha)$ in the point $O$ in the case $(b)$. In both cases, with known $|C_1|$ and $\theta_1$, we can use Eq. (\ref{phi via theta}) for finding the phase $\varphi_1$ of the parameter $C_1$ as a function of $B_+$. Similar measurements and calculations can be done with polarizers deviating for a small angle $\alpha$ from the vertical orientation, to determine $\theta_4$ and, then, $\varphi_4$ as a function of $B_+$. The last step is a single coincidence measurement with one of two polarizers turned for $45^\circ$ and the other one for $135^\circ$. This measurement gives $\left.w_{135^\circ}\right|_{45^\circ}$, and then Eqs. (\ref{Eq for B+})-(\ref{cos phi1-phi4}) can be used for finding $B_+$ and, then, $\varphi_{1,4}$. With $|B_-|$ found from Eq. (\ref{cond-HV-VH}),  this finalizes the procedure of finding all parameters of MPS of a general form.

\subsubsection{$B_+=0$}

In the picture $(c)$ of Fig. \ref{Fig2} the points $A$ and $B$ are located symmetrically with respect to the vertical axis crossing the point $O$. Again, three points $A$, $O$, and $B$ can be connected by a parabola,
\begin{equation}
 \label{parabola}
 \left.w_\alpha\right|_\alpha\approx\left.w_H\right|_H +k\alpha^2.
\end{equation}
The line tangent to this parabola in the point $O$ is horizontal. This means that in this case there is no validity region for the linear approximation of Eqs. (\ref{cos-phi1}) and (\ref{cos-phi4}), which is possible (at $C_{1,4}\neq 0$) only if $B_+=0$. In this case a direct measurement of the parameter $k$ in Eq. (\ref{parabola}) appears to be sufficient for completing the reconstruction of the parameters characterizing MPS. Indeed, in the case $B_+=0$ and $\alpha\ll 1$ Eq. (\ref{cos-phi1}) takes the form
\begin{equation}
 \label{C1 at zero B+}
 C_1^\alpha\approx C_1+ (C_4-C_1)\, \alpha^2.
\end{equation}
Moreover, as  $B_+=0$, we can choose now an arbitrary phase of $\Psi^{(3)}$ in a way providing $\varphi_1=0$, which leaves only one phase parameter $\varphi_4$ to be determined  form experiments in addition to $C_1$ and $C_4$. Under this assumption Eq. (\ref{C1 at zero B+}) gives
\begin{equation}
 \label{C1 squared}
 \left|C_1^\alpha\right|^2\approx |C_1|^2+ 2|C_1|(|C_4|\cos\varphi_4-|C_1|)\, \alpha^2
\end{equation}
and, being compared with Eq. (\ref{parabola}),
\begin{equation}
 \label{k via cos phi4}
 k=2|C_1|(|C_4|\cos\varphi_4-|C_1|).
\end{equation}
Thus, by measuring experimentally the parabola parameter $k$ of Fig. 2$(c)$ and Eq. (\ref{C1 at zero B+}) we find from Eq. (\ref{k via cos phi4}) the phase $\varphi_4$, and this concludes the reconstruction of all parameters of MPS in the case $B_+=0$.

\section{ Reconstruction of the ququart's parameters}

 Let us return now to pure states of biphoton ququarts. The analysis of the previous Section shows that  by means of purely polarization measurements one can determine all parameters of the ``qutrit's part" of the ququart $\Psi^{(3)}$ (except its phase) and the parameter $|B_-|$. The only remaining unknown parameter of the ququart's states is the phase $\varphi_-$ of the parameter $B_-$. Though characteristics of MPS do not depend of $\varphi_-$, features of pure states of ququarts can be phase-sensitive. The phase $\varphi_-$ cannot be found from any purely polarization measurements and requires combined polarization-frequency measurements. This means that in the experimental scheme of Fig. \ref{Fig2} one has to install in front of detectors both polarizers and frequency filters. Such coincidence measurements sufficient for determining the phase $\varphi_-$ are most simple in the case of ququarts with $B_+\neq 0$. Then, one of the measurable conditional probabilities is the probability of registering high-frequency horizontally polarized photons in the channel 1 under the condition that simultaneously one registers low-frequency vertically polarized photons in the channel 2
 \begin{gather}
  \nonumber
  \left.w_{H,h}\right|_{V,l}=\frac{\left.N_{H,h}\right|_{V,l}}{N_{tot}}=\frac{|B_++B_-|^2}{2}\\
  \label{cond-freq}
  =\frac{|B_+|^2+|B_-|^2+2|B_+||B_-|\cos(\varphi_--\varphi_+)}{2}.
 \end{gather}
By assuming again that $\varphi_+=0$, we find from this equation $\cos\varphi_-$ expressed in terms of the experimentally measurable relative amounts of photon counts.

The case $B_+=0$ (but $C_{1,4}\neq 0$) is not much more complicated or difficult. With polarizers in both channels 1 and 2 turned for an arbitrary but identical angle $\alpha$, one can use, in fact, the same scheme of measurements as described above for the case $B_+=0$. Indeed, as mentioned above, in the case $B_+=0$ Eq. (\ref{B+-alpha}) takes the form $B_+^\alpha=\sqrt{2}\cos\alpha\sin\alpha\, (C_4-C_1)$. As the parameters $C_1$ (with $\varphi_1=0$), and $C_4=|C_4|e^{i\varphi_4}$ are supposed to be known already from purely polarization measurements, we can write the difference $C_4-C_1$ as $C_4-C_1=|C_4-C_1|e^{i\varphi_{4-1}}$, where $|C_4-C_1|$ and $\varphi_{4-1}$ are easily calculable. Now, with the turned polarizers, one can measure the conditional probability of registering a high-frequency photon polarized in the direction $\alpha$ in the channel 1 and a low-frequency photon polarized in the direction $\alpha+90^\circ$ in the channel 2. This conditional probability is related to the unknown phase $\varphi_-$ by a formula very similar to that of Eq. (\ref{cond-freq})
\begin{gather}
  \nonumber
  \left.w_{\alpha,h}\right|_{\alpha+90^\circ,l}=\frac{\left.N_{\alpha,h}\right|_{\alpha+90^\circ,l}}{N_{tot}}
  =\frac{|B_+^\alpha +B_-|^2}{2}\\
  \label{cond-freq-alpha}
  =\frac{|B_+^\alpha|^2+|B_-|^2+2|B_+^\alpha||B_-|\cos(\varphi_--\varphi_{4-1})}{2},
\end{gather}
where $|B_+^\alpha|=\frac{1}{\sqrt{2}}|\sin 2\alpha||C_4-C_1|$.
Eq. (\ref{cond-freq-alpha}) can be used for finding the phase $\varphi_-$ from the data to be obtained from the coincidence polarization-frequency measurements in the case of ququarts with $B_+=0$.

\section{Conclusion}

Thus, biphoton ququarts are more complicated and their physics is more rich and interesting than usually assumed. The key elements of this newer understanding are (i) the obligatory symmetry of biphoton wave functions (in pure states) as wave functions of two indistinguishable bosons, and (ii) consideration of frequencies of photons in biphoton polarization-frequency ququarts as variables independent of polarizations rather than as given numbers. In this approach biphoton polarization-frequency ququarts are states having two degrees of freedom for each photon, polarization and frequency. Owing to this, all biphoton polarization-frequency ququarts are entangled and, in a general case, their entanglement is an inseparable mixture of the polarization and frequency entanglement. Another interesting consequence of the formulated features of biphoton ququarts concerns their images to be seen in experiments. If in fully polarization-frequency coincidence measurements ququarts are seen as pure states, in simpler purely polarization (non-selective in frequencies) measurements the same states are seen as two-qubit mixed polarization states, MPS. MPS are characterized by the ququart's density matrix reduced with respect to the frequency variables. Parameters of MPS are found. They appear are to be rather peculiar, differing significantly from those of the full-dimensionality ququarts, and rather useful. In particular, the Schmidt parameter of MPS, ${\overline K}$ is found to be related directly to the degree of polarization of ququarts. Features of MPS can be used straightforwardly for experimental measurement of ququart's parameters if they are not known in advance. A scheme of such measurements is suggested and described. The main idea is in separation of a biphoton beam for two channels by a simple non-selective beam splitter and in performing series of coincidence measurements with photon counters. The first stage consists in making purely polarization measurements with different orientations of polarizers in both channels and in finding in this way all parameters of MPS. Then for reconstructing completely the  ququart's state, one has to find additionally only one of its phase parameters. which requires making one additional polarization-frequency coincidence measurement with both polarizers and frequency filters installed in both channels in front of detectors. So, the scheme suggested for reconstruction of the ququart's parameters separates purely polarization and polarization-frequency measurements and minimizes the amount of polarization-frequency measurements. We believe that such experiments are feasible and their results may be sufficiently interesting.

\section*{Acknowledgement}
The work is supported partially by the grant RFBR 11-02-01043-a.

\end{document}